\documentclass{article}

\PassOptionsToPackage{numbers, compress}{natbib}
\usepackage[preprint]{neurips_2023}

\usepackage[utf8]{inputenc} % allow utf-8 input
\usepackage[T1]{fontenc}    % use 8-bit T1 fonts
\usepackage{hyperref}       % hyperlinks
\usepackage{url}            % simple URL typesetting
\usepackage{booktabs}       % professional-quality tables
\usepackage{amsfonts}       % blackboard math symbols
\usepackage{nicefrac}       % compact symbols for 1/2, etc.
\usepackage{microtype}      % microtypography
\usepackage{xcolor}         % colors
\usepackage{graphicx}
\usepackage{algorithm}
\usepackage{algorithmic}

\title{AF2-Mutation: Adversarial Sequence Mutations against AlphaFold2 on Protein Tertiary Structure Prediction}

\author{Zhongju Yuan\textsuperscript{1,5} \thanks{Work completed during an internship at Shanghai AI Lab by Zhongju Yuan.}, Tao Shen\textsuperscript{4}, Sheng Xu\textsuperscript{1}, Leiye Yu\textsuperscript{3}, Ruobing Ren\textsuperscript{3}, Siqi Sun\textsuperscript{2,1}\thanks{Corresponding author: siqisun@fudan.edu.cn, renruobing@fudan.edu.cn}\\
\textsuperscript{1}Shanghai AI Laboratory\\
  \textsuperscript{2}Research Institute of Intelligent Complex Systems, Fudan University\\
  \textsuperscript{3}Institute of Metabolism and integrative biology, Fudan University\\
  \textsuperscript{4}Zelixir Biotech\\
  \textsuperscript{5}WAVES, Ghent University\\
  \texttt{Zhongju.Yuan@UGent.be},
\texttt{\{yuleiye, renruobing, siqisun\}@fudan.edu.cn}\\
\texttt{shentao@zelixir.com},
\texttt{shengxu@link.cuhk.edu.hk}\\
}

\begin{document}

\maketitle
\begin{abstract}
Deep learning-based approaches, such as AlphaFold2 (AF2), have significantly advanced protein tertiary structure prediction, achieving results comparable to real biological experimental methods. While AF2 has shown limitations in predicting the effects of mutations, its robustness against sequence mutations remains to be determined. Starting with the wild-type (WT) sequence, we investigate adversarial sequences generated via an evolutionary approach, which AF2 predicts to be substantially different from WT. Our experiments on CASP14 reveal that by modifying merely three residues in the protein sequence using a combination of replacement, deletion, and insertion strategies, the alteration in AF2's predictions, as measured by the Local Distance Difference Test (lDDT), reaches 46.61. Moreover, when applied to a specific protein, SPNS2, our proposed algorithm successfully identifies biologically meaningful residues critical to protein structure determination and potentially indicates alternative conformations, thus significantly expediting the experimental process.
\end{abstract}

\section{Introduction}

Investigating tertiary protein structures offers insights into their functions, facilitating the development of innovative therapeutic interventions. Acquiring native protein structures\footnote{Experimentally determined ground-truth structure.} has long been a formidable challenge in biology. Recent advancements in deep learning and protein language models, particularly AlphaFold2 (AF2)~\cite{jumper2021highly}, have enabled more precise protein structure predictions. Notably, AF2 demonstrates outstanding performance across various protein families.

Although current deep-learning approaches have achieved remarkable accomplishments in diverse fields, adversarial samples often expose their limitations. Minor differences between adversarial and original samples can significantly impact the model's output. While attack strategies have been extensively explored in computer vision (CV)~\cite{elsayed2018adversarial, mahmood2021robustness,adesina2022adversarial} and natural language processing (NLP)~\cite{zhang2020adversarial, morris2020textattack} models, there is a research gap in the biological domain. % To the best of our knowledge, we are the first to investigate the impact of adversarial mutations on deep learning models such as AF2.

The discrete nature of protein sequences complicates adversarial attacks, rendering gradient-based techniques suboptimal for obtaining adversarial samples for discrete data. Moreover, AF2's "recycling" approach further complicates gradient computation. An evolutionary attack technique with gradient-free characteristics could simplify and enhance the process, compared to existing gradient-free attack methods that often require intricate computations.

\begin{figure*}[!t]

\begin{center}

\includegraphics[width=0.8\textwidth]{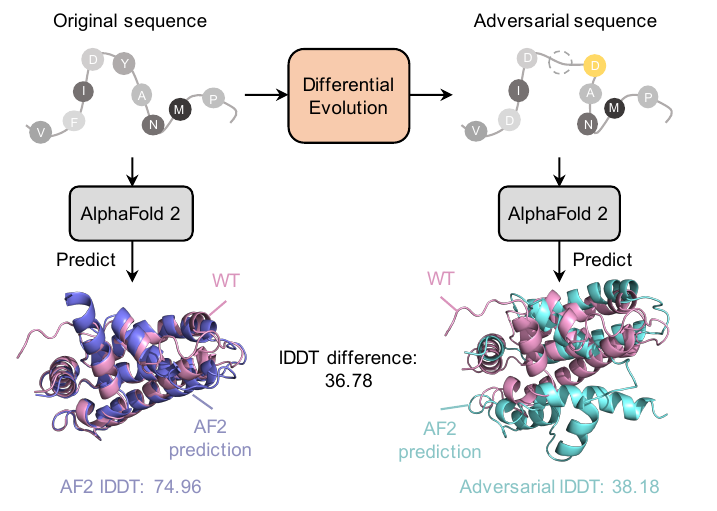}

\end{center}

\caption{\textbf{Procedure overview.} The Differential Evolution algorithm is employed to produce adversarial sequences for wild-type (WT) proteins, with the objective of maximizing the lDDT discrepancy between the predicted structure of the original and its mutated counterpart.}

\label{fig:motivation}

\end{figure*}

In this paper, we introduce an efficient method for generating adversarial mutation sequences for WT\footnote{The original protein sequences that are about to be mutated or attacked.}: \textbf{AF2-Mutation} (Figure~\ref{fig:motivation}). The core of the algorithm employs the differential evolution algorithm to identify the mutation strategies—including replacement, deletion, and insertion—that have the highest potential to affect the predicted structure. To preserve sequence similarity, mutations are restricted to three amino acids (AAs). We utilize the Local Distance Difference (lDDT) score~\cite{mariani2013lddt}, a superposition-free metric evaluating local distance differences of all protein atoms, to assess the similarity between the predicted protein structures and the native ones. Consequently, the lDDT gap before and after the attack serves as a measure of the attack's effectiveness.

Furthermore, we perform a case study on the natural protein sequence SPNS2 to verify the biological significance of mutations identified by AF2. Our method reveals that the proposed approach successfully detects several biologically relevant AAs. As the identified residues are crucial for determining protein structure and potentially suggesting alternative conformations, our method can substantially expedite the experimental process.

Our primary contributions include:

\begin{itemize}
    \item Proposing a novel framework for conducting adversarial attacks against AF2 using replacement and mixed strategies.
    \item Demonstrating the effectiveness of the framework on CASP14\footnote{\url{https://predictioncenter.org/casp14/}} targets by showing that some adversarial samples generated by missense mutations significantly mislead the AF2 model.
    \item Confirming the efficacy of the proposed method in detecting biologically meaningful alterations in real-world applications through experiments.
\end{itemize}

\section{Related Work}

Transformer-based deep learning models have been developed for predicting protein tertiary structures, with the recently introduced AF2 demonstrating promising outcomes. In addition to accurate protein structure prediction, assessing the robustness of AF2 to mutations is of interest. Researchers have generated several examples where protein sequences with differences in merely five residues lead to significantly distinct structures~\cite{jha2021protein}, verifying that another popular structure prediction model, RoseTTAFold~\cite{baek2021accurate}, lacks robustness. Nonetheless, efficient methods for attacking general proteins are lacking, with current research focusing on specific proteins and their alterations~\cite{alkhouri2023robustness}.

Presently, adversarial attack techniques are predominantly applied in CV and NLP. However, given that protein sequences constitute discrete data, unlike continuous data in CV, we concentrate on methods employed in NLP.

Experimental conditions can be classified into white box and black box categories, depending on the availability of gradients. The choice of attack positions and the characters or words to replace or insert vary due to the differences between these two categories. In a white-box setting, where the model's structure and parameters are known, gradient-based techniques can be employed to identify adversarial samples. These methods often aim to minimize a loss function while adhering to specific constraints~\cite{ebrahimi2018hotflip, zhao2017generating, michel2019evaluation, ren2019generating}. Conversely, in a black-box setting, the model's core details are not accessible. As gradients cannot be directly calculated, alternative metrics are used to rank the significance of characters or words in the sample, such as the output difference after removing a character or word~\cite{alzantot2018generating, jin2020bert, li2020bert}. Additionally, genetic algorithms have proven to be the most effective strategy for generating adversarial samples without using gradients~\cite{alzantot2018generating}. As a result, our paper employs differential evolution, an evolutionary algorithm, in a black-box setting.

Drawing parallels between attacking strategies on protein sequences and those employed in NLP techniques, three main strategies can be identified: replacement, insertion, and deletion. In the context of protein sequences, these approaches involve manipulating amino acids (AAs) instead of characters or words. The strategies focus on replacing AAs with similar or functionally related alternatives, which mirrors the replacement of characters with commonly misspelled options or semantically similar words in NLP~\cite{liang2018deep, hsieh2019robustness}. Recognizing the limitations of individual attacking strategies, researchers have sought to develop methods that combine these approaches, as exemplified in techniques like DeepWordBug~\cite{gao2018black}, HotFlip~\cite{ebrahimi2018hotflip}, and TextBugger~\cite{li2018textbugger, liang2018deep}.

\section{Methods}
\subsection{Problem Formulation}

Given a protein sequence of length $L$, denoted as $\mathbf{X} = \{r_{1}, \cdots, r_{i}, \cdots, r_{L}\}$, where $r_{i, i=1,\cdots, L}$ represents the $i$-th residue in the sequence, the native and predicted structure of the sequence are represented as $\mathbf{Y}$ and $\hat{\mathbf{Y}}$. A pre-trained model $F: \mathbf{X} \rightarrow \mathbf{Y}$, in this case, AF2, is utilized to predict the protein structure of the sequence. The accuracy of the predicted structure is evaluated by comparing it with the native structure using the lDDT metric. Adversarial attacks aim to generate an adversarial sample $\mathbf{X}_{mut}$ against the pre-trained model so that the objective function in Eq.~\ref{eq:objective} can be minimized.
\begin{equation}
\footnotesize \mathrm{lDDT}(F(\mathbf{X}_{mut}), \mathbf{Y})-\mathrm{lDDT}(\hat{\mathbf{Y}}, \mathbf{Y}).
\label{eq:objective}
\end{equation}

\begin{figure*}[!t]

\begin{center}

\includegraphics[width=0.9\linewidth,trim=0 0 0 0,clip]{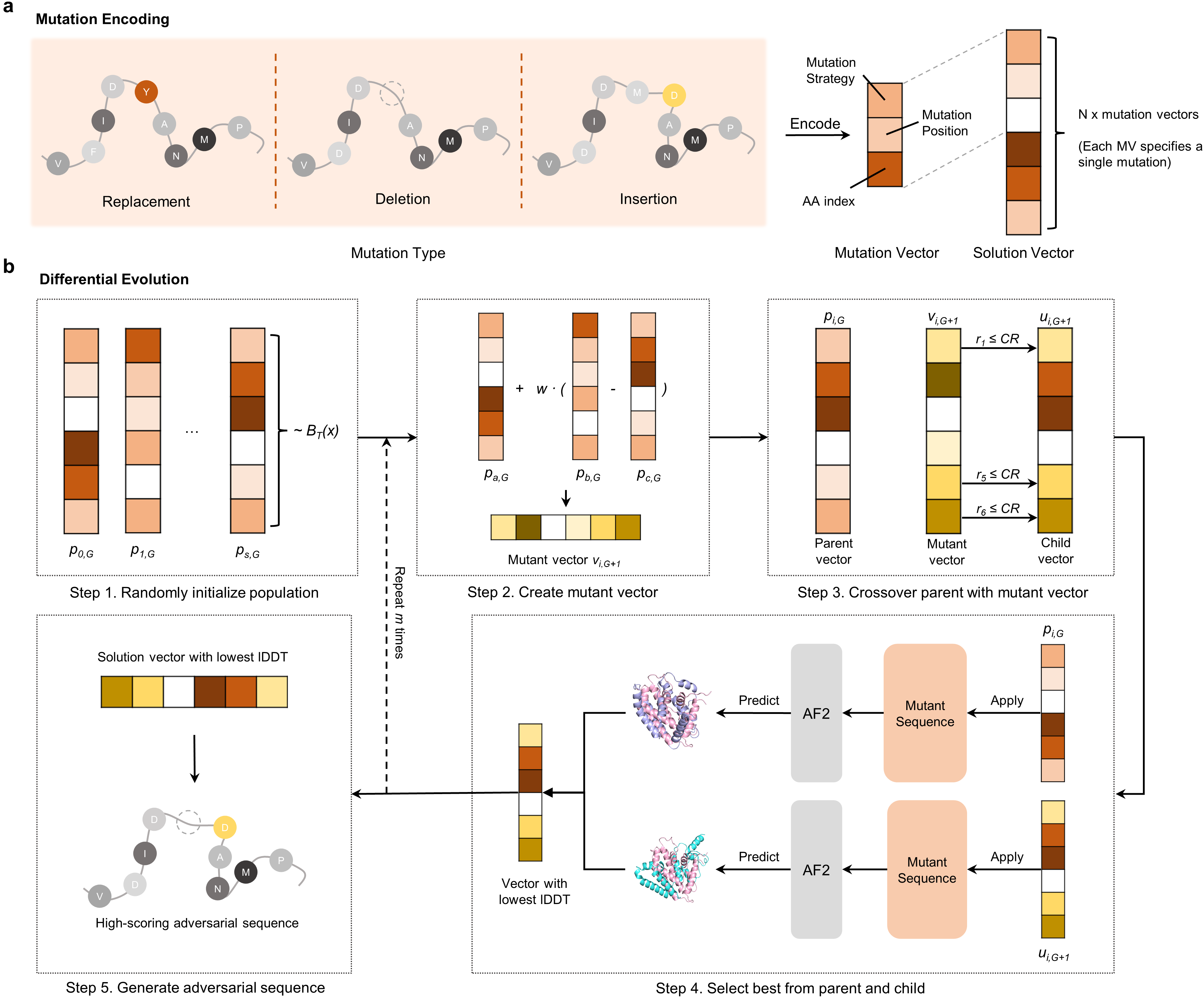}

\end{center}

\caption{The overview of AF2-mutation. (a) Mutation Encoding: This illustration demonstrates how a solution vector is used to encode an attack method. (b) Differential Evolution: This graphic depicts the iterative process leading to the solution vector that results in the ultimate adversarial sample.}

\label{fig:overview}

\end{figure*}

The basis of our method for evaluating AF2's robustness is the idea that an adversarial sample might produce a structure significantly different from the native one. Figure \ref{fig:overview} illustrates the general overview of our method.

\subsection{AF2-Mutation Algorithm}
In this section, we cast the attack strategy as an optimization problem, aiming to select optimal amino acid (AA) locations to maximize the local Distance Difference Test (lDDT) value between adversarial and native structures. This optimization problem has been defined, and we provide an optimal solution vector.

Due to the substantial search space of this optimization problem, exhaustive search is unfeasible, and random search is unlikely to provide a globally optimal result, or even a locally optimal one. To tackle this challenge, we propose the differential evolution method, a gradient-free, population-based evolutionary optimization approach. This method identifies the optimal solution without requiring the gradient of the AA embedding.

We construct the search space considering three distinct types of attack strategies against AF2 (as shown in Figure~\ref{fig:overview}(a) left):
\begin{itemize}
\item Replacement: Substituting AAs in the protein sequence with one of the remaining 19 AAs.
\item Deletion: Removing a residue in the protein sequence.
\item Insertion: Selecting one of the 20 AAs for insertion into the protein sequence.
\end{itemize}

In attacking a sequence, we employ these three strategies together and encode the mutation method into a solution vector of length $N=3b$. Here, $3$ denotes the length of a mutation vector for each position (illustrated in Figure~\ref{fig:overview}(a) right). Each mutation vector consists of three components: the mutation strategy ($\mathrm{ind}$), the mutation site (or the index of the AA in the wild-type (WT) sequence), and the AA used for replacement and insertion. We set a budget of $b$, limiting the total number of mutated AAs in the sequence to it.

\begin{equation}
\begin{small}
\mathrm{Chosen \, strategy}=\left\{
\begin{array}{ll}
\mathrm{Replacement} & 0<\mathrm{ind}<0.33\\
\mathrm{Deletion} & 0.33 \leq \mathrm{ind} < 0.66\\
\mathrm{Insertion} & {\rm Otherwise}
\end{array} \right.. \\
\label{eq:index}
\end{small}
\end{equation}

The method inputs a protein sequence $\mathrm{X}$, the number of evolution iterations $m$, the population size $s$,  the crossover probability $\mathrm{CR}$, and the differential weight $W$ - parameters standard in differential evolution~\cite{storn1997differential}. The core idea of the method is to create a population of candidate solutions and iteratively improve these candidates until an optimal or satisfactory solution is found.

Figure \ref{fig:overview}(b) displays the generation of an initial set of candidate solution vectors $P={p_{0, G},\cdots,p_{i, G},\cdots,p_{s, G}}$ in Step 1, where each $p_{i,i=0,\cdots,s}$ is randomly selected, and $G$ indicates the generation index. This is followed by iterative enhancements until the best solution is discovered or the maximum iteration number $m$ is reached.

During each iteration, the differential evolution method generates new solution vectors using mutation, crossover, and selection operators. Step 2 in Figure~\ref{fig:overview}(b) shows the mutation operator generating new solution vectors as follows:
\begin{equation}
{v}_{i, G+1} = p_{a, G}+W*(p_{b, G}-p_{c, G}),
\end{equation}
where $v_{i, G+1}$ is the solution vector; $p_{a, G}$, $p_{b, G}$, and $p_{c, G}$ are distinct solution vectors randomly selected from the previous generation, and their lengths are all $|v_{i, G+1}|$, equal to $3b$.

The crossover operator, depicted in Step 3 of Figure~\ref{fig:overview}(b), fosters the diversity of the solution vectors. Following mutation in each generation, the child vector $u_{i, G+1}$ is generated:
\begin{equation}
u_{i_k, G+1}=\left\{
\begin{array}{rcl}
{v}_{i_{k}, G+1} & & if \, r_{i} \leq \mathrm{CR} \, or \, R=k\\
p_{i_{k}, G} & & if \, r_{i} > \mathrm{CR} \, and \, R \neq k
\end{array} \right., \\
\end{equation}
where $k, k \in {1, \cdots, 3b}$ is the k-th value in a vector, $r_{i}$ is the i-th value randomly generated from a uniform distribution, $\mathrm{CR}$ is the crossover rate, and $R$ is a randomly chosen integer from $1$ to $3b$.

The selection process for the next generation is illustrated in Step 4 of Figure~\ref{fig:overview}(b). The newly generated child solution vector $u_{i, G+1}$ is compared with the parent solution vector $p_{i, G}$. If the child vector has a lower objective function value, it will replace the parent as the solution vector; otherwise, the original solution value is retained.

In the final step (see Figure~\ref{fig:overview}(b)), the proposed strategy uses the lDDT between the predicted and native protein structures as the objective function (Eq.\ref{eq:objective}). The aim of differential evolution is to maximize the objective function. When the algorithm converges or reaches the preset maximum number of iterations, the adversarial missense mutation sample and the objective function value are obtained.

\subsection{MSA Approximation during Evolutionary Process}

\begin{figure}
\centering
\includegraphics[width=0.9\linewidth]{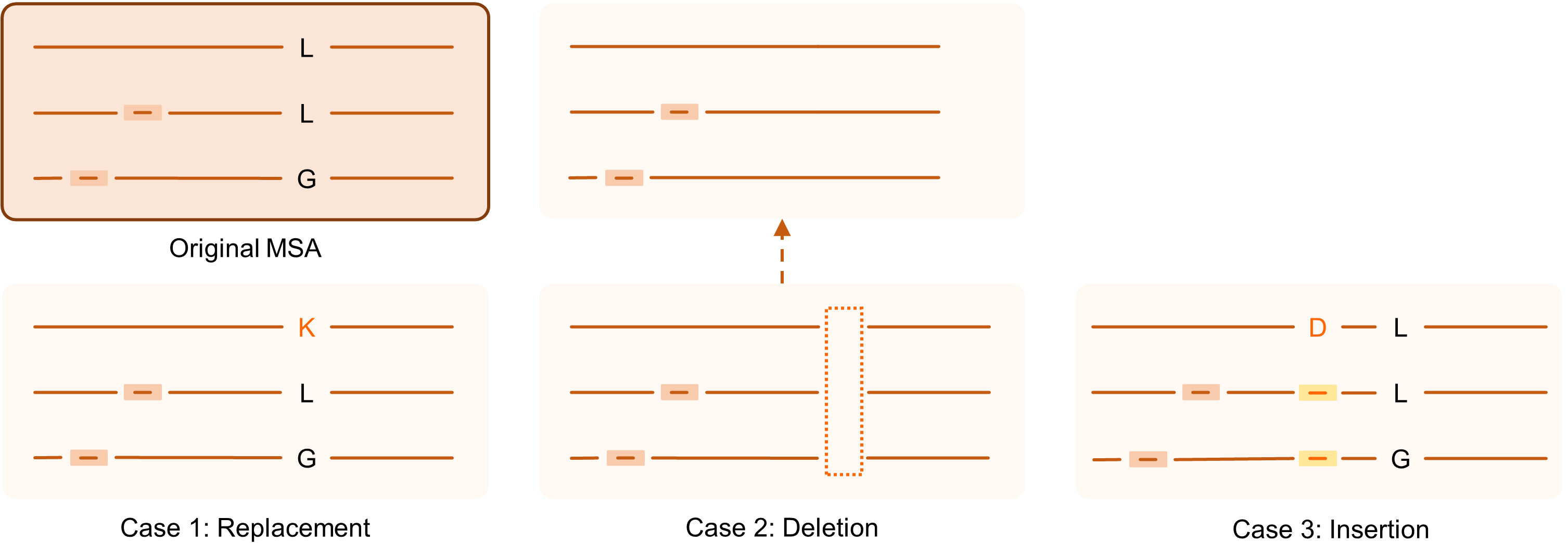}
\caption{Visualization of the proposed approximation algorithm to circumvent the need for realignment of MSAs.}
\label{fig:MSA_change}
\end{figure}

The inclusion of co-evolutionary information within MSAs is crucial for the accurate prediction of protein structures by AF2. However, identifying homologs in MSAs from the sequence database is a highly time-consuming process due to the vast number of sequences. Consequently, rebuilding MSAs for every mutated sequence within the evolutionary algorithm is impractical, necessitating the development of an approximate algorithm. To preserve the sequence length consistency between the MSA and the adversarial example, while also minimizing the impact of mutations on the MSA, we propose an approximation for each of the replacements, insertions, and deletions. As illustrated in Figure~\ref{fig:MSA_change}, the approximations can be grouped into three categories:
\begin{itemize}
\item Replacement (case 1): Following amino acid (AA) replacement, homologs of the mutated sequence remain unchanged, as the mutation does not affect the alignment at other locations.
\item Insertion (case 2): Upon the addition of a new residue to the query sequence, a gap character is inserted at the corresponding position in the other sequences within the MSAs. Consequently, alignment quality is preserved for the remaining sequences.
\item Deletion (case 3): When AAs are removed from a protein sequence, the corresponding positions in the other sequences within the MSAs are also eliminated.
\end{itemize}

To maximize the accuracy of AF2 predictions, we re-align the MSAs when predicting the structure of the final adversarial sample.

\section{Results}
\subsection{Experimental Setup and Dataset}
In this study, we evaluate the efficacy of our proposed replacement and mixed attack strategies against AF2 in a black-box setting for protein structure prediction. The experiments are conducted on the CASP14 dataset and a case study involving SPNS2. The Local Distance Difference Test (lDDT) serves as the metric to assess the similarity between two protein structures. 
%lDDT is a superposition-free score that evaluates local distance differences of all atoms in a protein, including the validation of stereochemical plausibility~\cite{mariani2013lddt}.

For each attack type, a differential evolution algorithm is employed to find the optimal solution. The population size and the number of evolution iterations are set to $p=32$ and $m=4$, respectively. Although increasing these parameters may yield improved solutions, it would also result in increased computation time.

To minimize the influence of the template on the final results, we utilize model 3 from the OpenFold~\cite{Ahdritz2022.11.20.517210} implementation of AF2 in our experiments, which omits template information. Furthermore, random masks and dropouts are set to zero to eliminate any potential impact on the experimental outcomes.

MSAs are constructed using the JackHMMER algorithm~\cite{johnson2010hidden} to iteratively search for candidate sequences in MGnify~\cite{mitchell2020mgnify} and UniRef90~\cite{suzek2007uniref}, and HHBlits~\cite{remmert2012hhblits} to search in UniClust30~\cite{mirdita2017uniclust} and the Big Fantastic Database (BFD)\footnote{\url{https://bfd.mmseqs.com}}. Candidate sequences are then aligned to the query sequence. As MSA construction is time-consuming, MSAs are generated only during the initial prediction of the WT sequence and during the final evaluation to obtain an accurate folding result.

% \subsection{Attack Results on CASP14}

\subsection{Attacking Results on CASP14}

\begin{figure}[!t]
\centering
\begin{minipage}{0.45\linewidth}
\includegraphics[width=\linewidth]{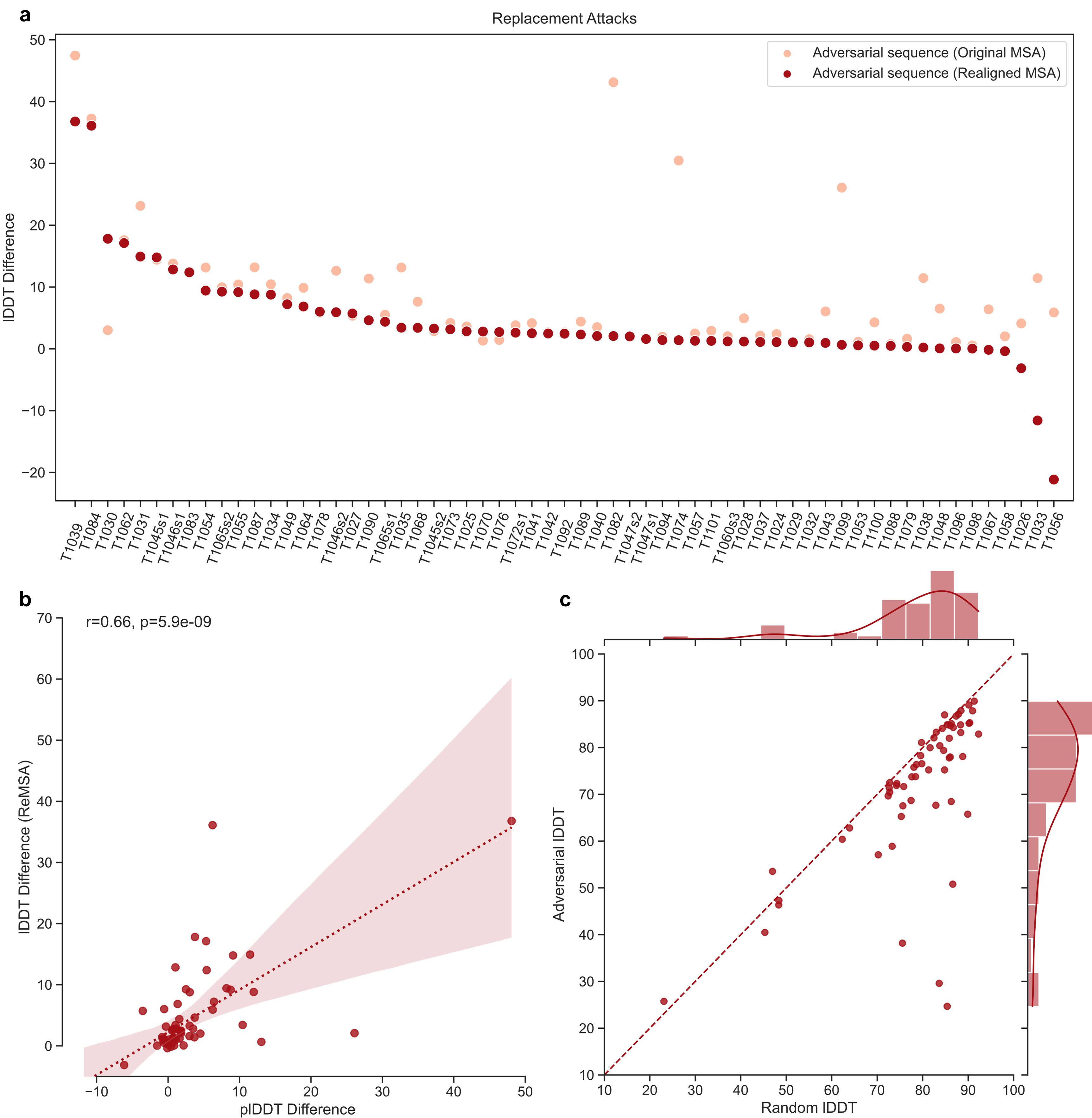}
\end{minipage}
\hfill
\begin{minipage}{0.47\linewidth}
\includegraphics[width=\linewidth]{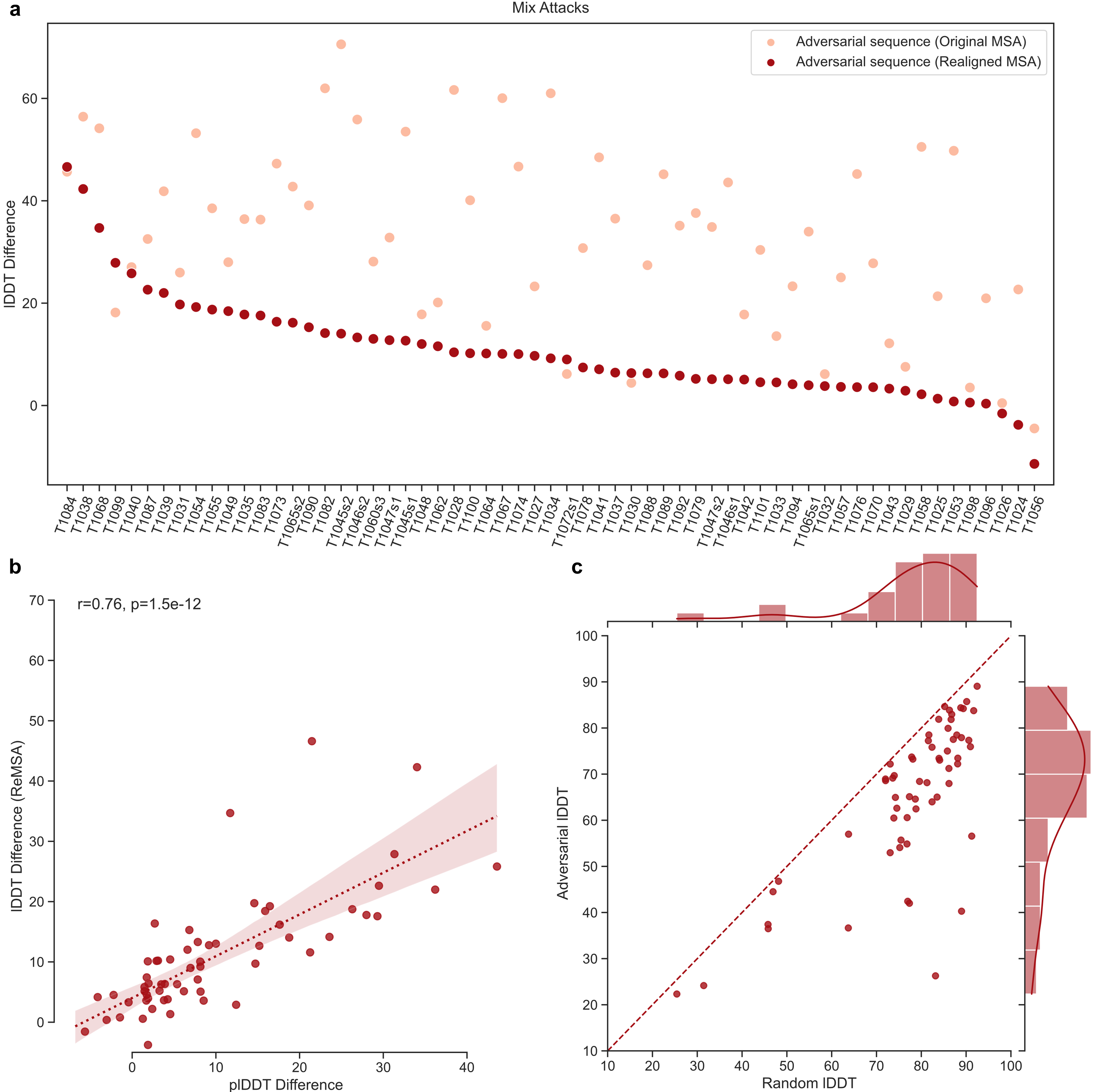}
\end{minipage}
\caption{Experimental results for the \textbf{replacement attack} (left) and the \textbf{mixed attack} (right) are presented as follows: (a) The figures depict the lDDT difference between the native structure and the adversarial structure, as well as the difference before and after re-alignment. (b) The figures illustrate the correlation between plDDT and lDDT. (c) The figures compare the proposed replacement and mixed attack methods with a random attack, and also display the distribution of these two results.}
\label{fig:rep_mix_res_fig}
\end{figure}

\begin{figure*}[!t]
\begin{center}
\includegraphics[width=1.1\linewidth, trim=50 0 0 0, clip]{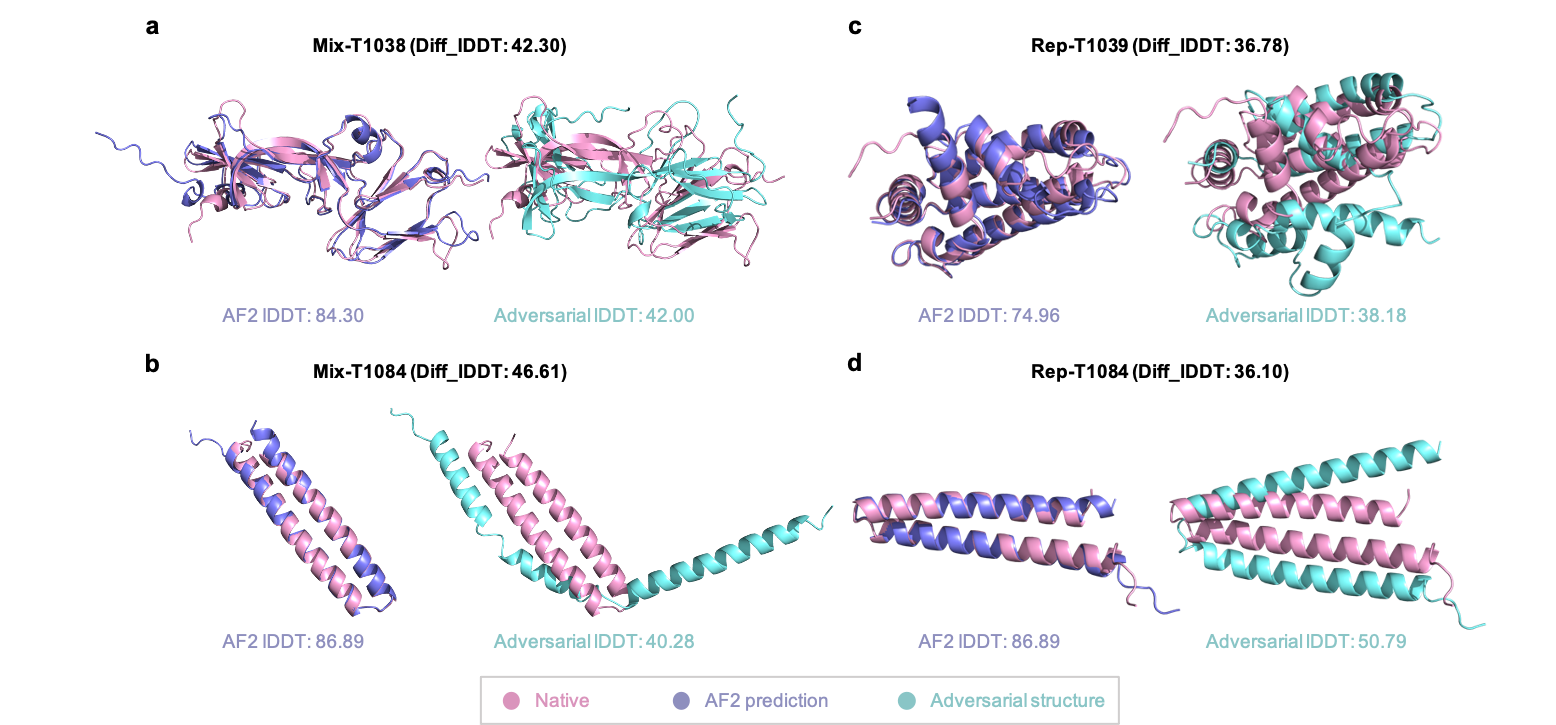}
\end{center}
\caption{Comparisons between the native structures and the AF2 predicted ones, including the original predicted structure and the prediction after the attack.}
\label{fig:res_structure}
\end{figure*}

All of the proteins in the CASP14 dataset are subjected to adversarial attacks for this paper. In conclusion, we are able to create adversarial sequences with, on average, more than 4, and 10 lDDT changes by replacement attacking and mixed attacking three AAs into the original WT sequences, indicating a substantial folding shift in 3D space.

Figure~\ref{fig:rep_mix_res_fig} (left) and (right) show the experimental results for the replacement and mixed attacks, respectively. In (a), the figures depict the lDDT difference between the native structure and the adversarial structure, as well as the difference before and after re-alignment. Comparing AF2's lDDT values for both the adversarial sample's prediction structure and the original sequence demonstrates that the proposed mixed attack algorithms are highly effective, with over half of the sequences showing differences of more than 10 lDDT (dark maroon dots). In particular, the mixed attack method (-10.76 lDDT on average) is consistently more effective than the replacement approach (-4.4 lDDT on average), which is not surprising since the mixed attack method searches over a broader space. It is interesting to note that there are two examples of replacement attacks that result in more accurate predictions than WT, and we will investigate them further in the future.

In Figure~\ref{fig:rep_mix_res_fig}(b), the figures illustrate the correlation between plDDT and lDDT for both replacement and mixed attacks. In (c), the figures compare the proposed replacement and mixed attack methods with a random attack, and also display the distribution of these two results. The proposed method is superior to the random attack in both cases. Almost all adversarial samples generated by our method outperform those produced randomly. In particular, the mixed attack drops 12.33 more than the random attack on average, while the replacement attack loses 7.03 more.

\subsection{Relation to plDDT and the Random Attack Baseline}

Predicted lDDT (plDDT) is an output produced by AF2 that mimics lDDT without knowing the native (ground-truth) structure, which also measures how confident the algorithm is in its predictions. Hence, mutating sequences will also alter plDDT since the input has changed. Figure~\ref{fig:rep_mix_res_fig}(b) demonstrates a positive correlation between plDDT and lDDT as determined by replacement (left) and mixed (right) attacks, which are 0.66 and 0.76, respectively. Nevertheless, the plDDTs for the two strategies differ significantly: the values of plDDT for the replace strategy (left) are close to zero for the majority of points, while those for the mixed method (right) are spread over the x-axis. AF2 appears to be more confident in replacing AAs, possibly due to its masked language model objective during its training and inference phase.

Furthermore, random attacks are used as baselines to evaluate the proposed approach's effectiveness. As part of the baseline algorithm, the positions to attack and residues to replace are randomly chosen. In order to establish a fair comparison, the total number of iterations for the differential evolution method and the number of random attacks are the same. As a result of a series of such random attacks, the mutation with the lowest lDDT value is retained as the result of the baseline. Comparison of the proposed mixed attack and the random attack is shown in Figure~\ref{fig:rep_mix_res_fig}(c), revealing the superiority of the proposed method over the random attack. Almost all adversarial samples generated by our method (left for replacement and right for mixed attacks) outperform those produced randomly. In particular, the mixed attack drops 12.33 more than the random attack on average, while the replacement attack loses 7.03 more.

\subsection{Necessity of Re-Alignment}
As mentioned earlier, the final result is computed by using newly aligned MSAs in order to provide a more accurate estimate of AF2. In this section, we demonstrate the need for such re-alignment prior to the evaluation.

In an attempt to reduce the complexity of the attacking algorithm, we propose an approximation algorithm that mimicked the new MSAs after the mutation, rather than searching MSAs from a vast sequence database. It may seem intuitive that changing one AA and such MSAs should not result in a significant difference between attacks. Still, the approximation error could accumulate after each mutation, eventually leading to a dramatic difference. Figure~\ref{fig:rep_mix_res_fig}(a) demonstrates the difference in the lDDT between approximated and re-aligned MSAs for replacement (left) and mixed (right) attack algorithms, respectively. For the mixed attack algorithm, a substantial difference is observed between the predicted structure and native structure for the approximated MSA (light maroon dots, -33.28 lDDT) as well as the re-aligned ones (dark maroon dots, -10.76 lDDT), indicating such approximation error accrues over time, and it is imperative that the MSA be re-aligned to obtain an accurate assessment. Similar results are observed for replace attacks as well, where the approximated MSA and re-aligned MSA result in -8.66 and -4.42 lDDT, respectively.

\begin{figure}[!t]
\begin{minipage}{0.35\textwidth}
\centering
\includegraphics[width=0.9\linewidth]{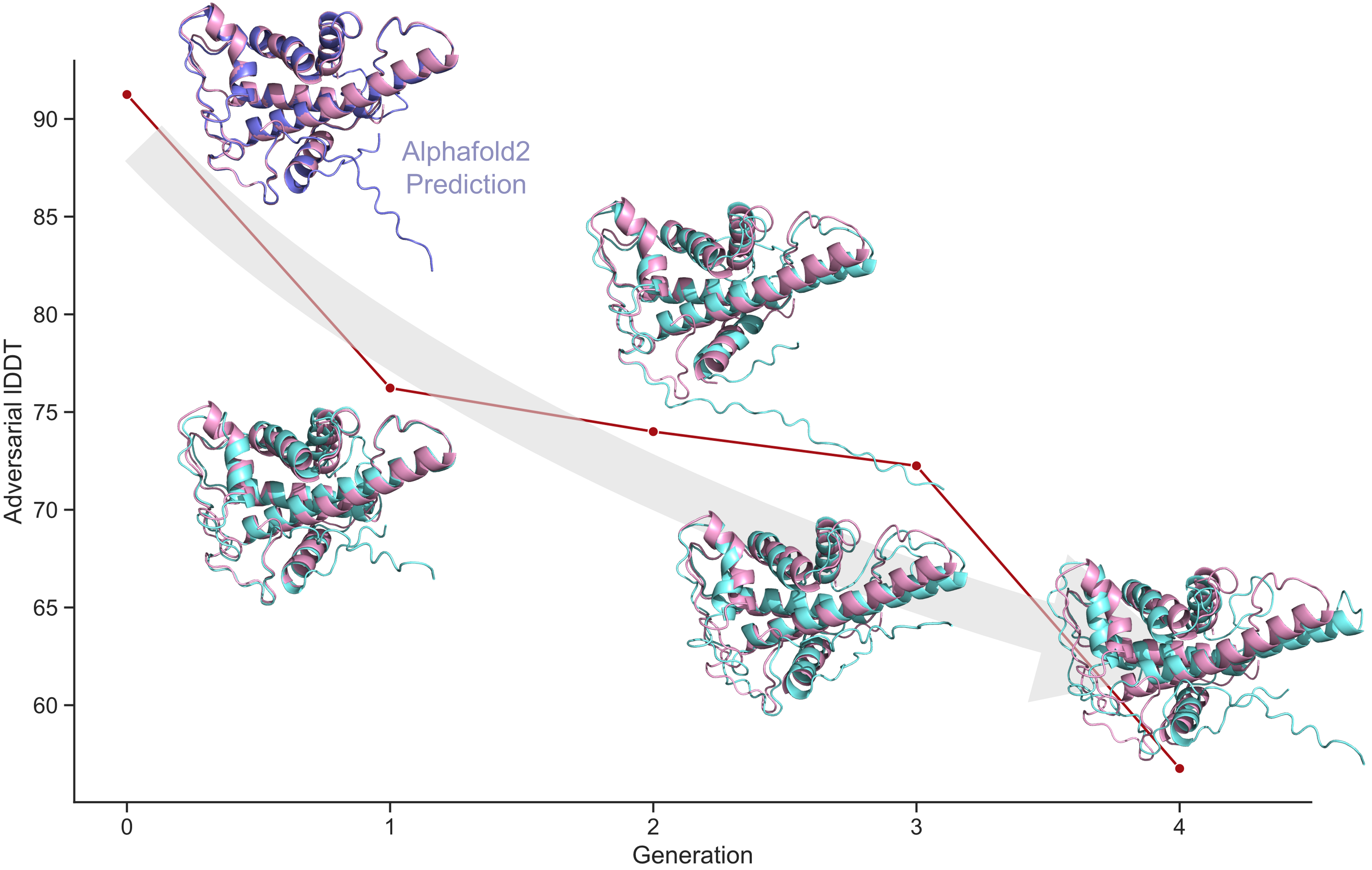}
\caption{The evolution process for the mixed attack on T1068.}
\label{fig:change_T1068}
\end{minipage}
\hfill
\begin{minipage}{0.6\textwidth}
\centering
\includegraphics[width=0.65\linewidth]{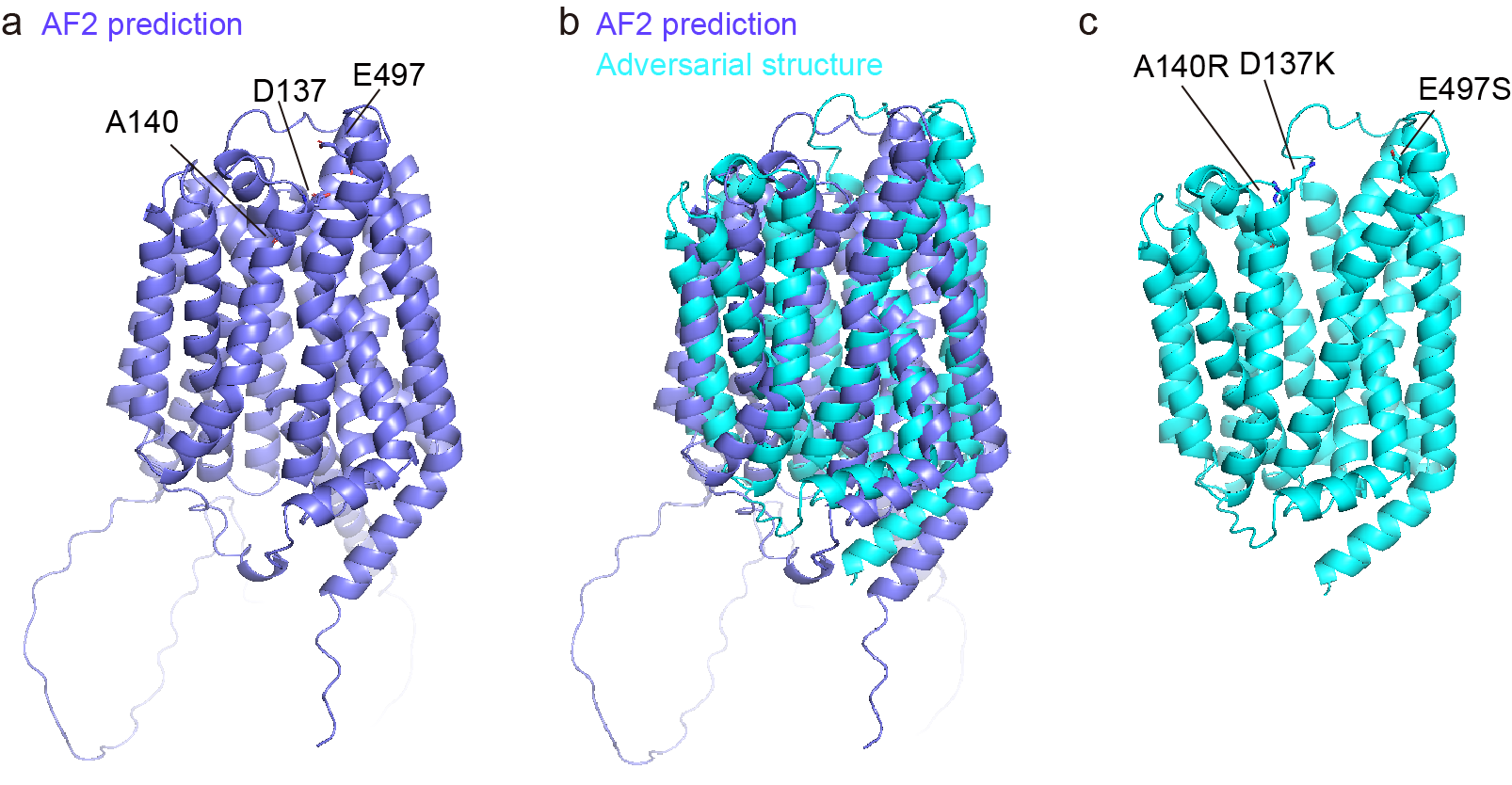}
\caption{Conformational change of the predicted structure of SPNS2 after the attack.}
\label{fig:real_pro_fig}
\end{minipage}
\end{figure}

% \begin{figure}[!t]
% \begin{center}
% \includegraphics[width=0.8\linewidth]{figs/change_process.png}
% \end{center}
% \caption{The evolution process for the mixed attack on T1068.}
% \label{fig:change_T1068}
% \end{figure}

% \begin{figure}[!t]
% \begin{center}
% \includegraphics[width=0.8\linewidth]{figs/Membrane_protein.png}
% \end{center}
% \caption{Conformational change of the predicted structure of SPNS2 after the attack.}
% \label{fig:real_pro_fig}
% \end{figure}

\subsection{Case Study}

We selected the two most effective targets in both replacement and mixed attack scenarios, and the detailed results are presented in Figure~\ref{fig:res_structure}. In both effective targets in the mixed attack scenarios (Figure~\ref{fig:res_structure}(a) and (b)), all main chains of the two targets have a substantial deflection or rotation. Whereas in both effective targets in the replacement attack scenarios (Figure~\ref{fig:res_structure}(c) and (d)), noticeable conformational change also occurs.

To illustrate the variations of prediction in the evolution process, we used a 4-generation evolutionary process of T1068, a sample with reasonably satisfactory AF2 prediction accuracy (91.25 lDDT). As a result of the three iterations, the lDDT decreased by 15, 17, and 19 points, respectively. While in the final generation, it is further reduced by 56.76 points. Despite re-alignment, the difference remains at 34.69. The visualization in Figure~\ref{fig:change_T1068} illustrates that each iteration produced a significant conformation change, indicating the proposed algorithm's effectiveness and approximated MSA approach.

Finally, we use a major facilitator superfamily(MFS) protein, SPNS2, for testing to further verify that the algorithm's detected location is of biological significance. The MFS is the largest secondary active transporter superfamily. It is a large class of integrated membrane proteins that mediate the transmembrane transport of various small molecules across the plasma membrane and between organelles. The structure model of SPNS2 predicted by Alphafold2 is an inward open conformation (Figure~\ref{fig:real_pro_fig}(a)). The predicted perturbed positions are 41, 44, and 391, which correspond to D137, A140, and E497 of the original sequence, respectively (Figure~\ref{fig:real_pro_fig}(a)). The perturbed positions (D137, A140, and E497) are replaced with lysine (K), arginine (R), and serine (S), respectively. The structure model predicted after these mutations is an outward open conformation (Figure~\ref{fig:real_pro_fig}(b)-(c)). Compared with the inward open structure predicted by Alphafold2, the intracellular opening is closed, and the extracellular opening is formed in the predicted mutation structure. A previous study reported that D137 and R342 form a salt bridge on the extracellular side of SPNS2, which may be responsible for stabilizing the inward open conformation~\cite{dastvan2022proton}. E497 may also interact with R342 on the side of R342, which may play a role in stabilizing the extracellular close state of the inward open conformation. The mutation of D137 K may affect the salt bridge formation between D137 and R342. The mutation of D137K, A140R, and E497S may help the formation of the outward open conformation. Transport ligands by the MFS superfamily, including SPNS2, involves transforming between the inward and outward open conformations. Predicting the different conformations of SPNS2 will significantly promote the explanation of its functional mechanism.

\section{Conclusion}

In recent years, significant progress has been made in protein structure prediction. In this study, we investigated black-box gradient-free adversarial attacks on the recently proposed AF2 protein folding language model. Our approach proved successful in generating missense mutation adversarial samples, as evidenced by extensive experimental results. It was observed that the predicted structure of certain protein sequences could be dramatically altered by modifying a mere three residues, leading to a sharp decrease in lDDT. This finding highlights the vulnerability of AF2 to attacks involving three residues. Moreover, our proposed method is capable of predicting protein conformational changes and identifying critical locations for these changes.

% In recent years, the prediction of protein structure has advanced significantly. In this paper, we looked into black-box gradient-free adversarial attacks on the recently proposed AF2 protein folding language model. Our approach was effective at generating missense mutation adversarial samples, according to comprehensive experimental results. The predicted structure of some protein sequences can be drastically altered by changing just 3 residues since the lDDT drops precipitously. This demonstrated that AF2 is vulnerable to attacks involving three residues. The proposed method can also predict protein conformational changes and pinpoint key locations for such changes.

% \section{Supplementary Material}

% Authors may wish to optionally include extra information (complete proofs, additional experiments and plots) in the appendix. All such materials should be part of the supplemental material (submitted separately) and should NOT be included in the main submission.

% \section*{References}

\bibliographystyle{unsrt}  
\bibliography{reference}

\end{document}